# Non-volatile voltage control of magnetization and magnetic domain walls in magnetostrictive epitaxial thin films.

D.E. Parkes[1], S.A. Cavill[2], A.T. Hindmarch[1,a)], P. Wadley[1], F. McGee[1], C.R. Staddon[1], K.W. Edmonds[1], R.P. Campion[1], B.L. Gallagher[1] and A.W. Rushforth[1*]

[1]*School of Physics and Astronomy, University of Nottingham, Nottingham NG7 2RD, United Kingdom.*

[2]*Diamond Light Source Chilton, Didcot, Oxfordshire OX11 0DE UK.*

[a)] *Present address: Centre for Materials Physics, Department of Physics, Durham University, South Road, Durham, DH1 3LE, United Kingdom*

*\* Corresponding author*

*We demonstrate reproducible voltage induced non-volatile switching of the magnetization in an epitaxial thin $Fe_{81}Ga_{19}$ film. Switching is induced at room temperature and without the aid of an external magnetic field. This is achieved by the modification of the magnetic anisotropy by mechanical strain induced by a piezoelectric transducer attached to the layer. Epitaxial $Fe_{81}Ga_{19}$ is shown to possess the favourable combination of cubic magnetic anisotropy and large magnetostriction necessary to achieve this functionality with experimentally accessible levels of strain. The switching of the magnetization proceeds by the motion of magnetic domain walls, also controlled by the voltage induced strain.*

Pacs numbers. 75.50.Bb, 75.47._m, 75.60.-d, 75.70.Ak, , 75.80.+q

The control of the magnetization of a ferromagnetic layer by an electric field is a desirable objective for the implementation of magnetic materials in information storage [1,2] and logical processing devices [3]. The approaches adopted to achieve this aim include direct electric field gating [4] to change the anisotropy of an ultrathin ferromagnetic layer, and inducing mechanical strain to change the anisotropy of a magnetostrictive ferromagnetic layer in a hybrid ferromagnet/piezoelectric device [5, 6]. The latter approach has been presented in a concept for the design of high performance voltage controlled magnetic random access memory (MRAM) [7]. The implementation of this technique in non-volatile information storage devices will require the ability to induce two different magnetic easy axes in the ferromagnetic layer when the piezoelectric layer is in the zero voltage state. Recently, Wu et al [8] demonstrated that this can be achieved by inducing two different remnant strain states in a single crystal piezoelectric $[Pb(Mg_{1/3}Nb_{2/3})O_3]_{(1-x)}$-$[PbTiO_3]_x$ (PMN-PT) layer coupled to a Ni layer. An alternative method involves using a ferromagnetic layer with a cubic magnetocrystalline anisotropy to define two easy axes separated by 90° and then using the anisotropy induced by the mechanical strain to overcome the hard axis barrier between the magnetic easy axes to switch the direction of the magnetization. The second technique has the advantage that the implementation in nanoscale devices will not require engineering and controlling the position of ferroelectric domain walls, a field of study that is less advanced than the study of domain walls in ferromagnetic materials. However, switching of the magnetization by this method has not yet been achieved without the aid of an external magnetic field [9]. This is due to the difficulty in identifying a material with the favourable combination of a cubic magnetic anisotropy and a sufficiently large magnetostriction to overcome the anisotropy energy for experimentally achievable levels of strain. In this letter we show that a single crystal thin film of Galfenol, an alloy of Fe and Ga, possesses the required properties to achieve non-volatile switching of the magnetization direction at room temperature in the absence of a magnetic field. This demonstration is achieved by the application of a modest uniaxial strain of order $1 \times 10^{-4}$.

Alloying Fe with Ga has been shown to enhance the magnetostriction by over an order of magnitude in bulk samples for Ga concentrations in the range 17-27% [10]. Typical values of the magnetostriction constant, $3/2\lambda_{100} \approx 4 \times 10^{-4}$, are an order of magnitude higher than those of most ferromagnetic transition metals and alloys and are rivalled only by rare-earth compounds, such as Terfenol-D, used in many commercial applications [11]. Initial studies of the magnetostriction in $Fe_{1-x}Ga_x$ alloys focused on the behavior of bulk quench cooled samples [10]. Recently, epitaxial thin films of $Fe_{1-x}Ga_x$ have been grown by molecular beam epitaxy (MBE) [12,13]. Such films have been shown to possess a cubic magnetocrystalline anisotropy, but the magnetostriction was not studied directly due to clamping of the film to the substrate. In this Letter we exploit the inverse magnetostriction (or Villari) effect in a hybrid ferromagnet/piezoelectric device to demonstrate that a thin epitaxial $Fe_{81}Ga_{19}$ film grown on a GaAs substrate does indeed possess a high magnetostriction [14]. The combination of large magnetostriction and cubic magnetocrystalline anisotropy allows the demonstration of voltage controlled non-volatile switching of the magnetization direction in the absence of a magnetic field. The switching proceeds through the relatively large scale motion of magnetic domain walls between pinning sites, controlled by the voltage induced strain.

A 22.5nm $Fe_{81}Ga_{19}$ single crystal layer was grown by low temperature (0°C) MBE onto a GaAs(001) substrate. A 5nm amorphous GaAs capping layer was grown to protect the metallic layer from oxidation. X-ray diffraction measurements (Fig. 1(a)) using a Philips X-Pert materials research diffractometer show a single peak corresponding to the $Fe_{81}Ga_{19}$ layer, indicating that the layer is a single crystal phase with a

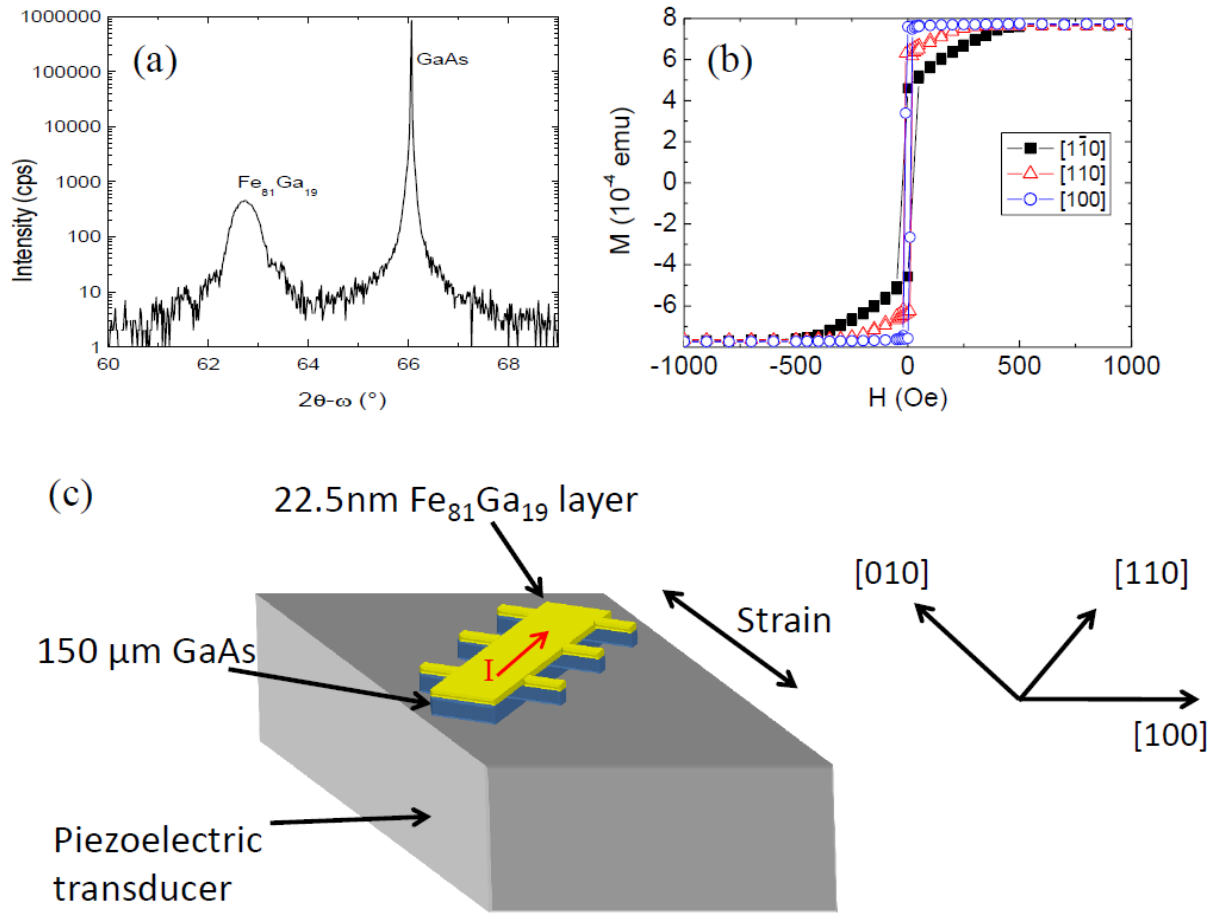

**Figure 1** (color online) (a) X-ray diffraction 2θ-ω scans of the $Fe_{81}Ga_{19}$ film on a GaAs substrate showing the sharp reflection from the GaAs substrate and the $Fe_{81}Ga_{19}$ layer peak. (b) Magnetic hysteresis loops of the unprocessed $Fe_{81}Ga_{19}$ sample measured at 300K by SQUID magnetometry. (c) A schematic diagram of the Hall bar device. The crystal directions of the GaAs and FeGa film are shown. The piezoelectric transducer is polycrystalline.

vertical lattice parameter of 0.296nm. The full width at half maximum of the $Fe_{81}Ga_{19}$ peak corresponds to a coherent crystal structure over the full 22.5nm thickness of the layer, and the epitaxial growth was confirmed by the reflection high energy electron diffraction (RHEED) pattern observed during the growth (not shown). Magnetic hysteresis loops measured on an unprocessed sample using a Quantum Design superconducting quantum interference device (SQUID) magnetometer (Fig. 1(b)) show that the layer has the magnetic easy axis in the plane. Fits to the data [15] reveal that the in plane magnetic anisotropy consists of a superposition of a strong cubic term ($K_C$=32.7kJm$^{-3}$) favouring the [100]/[010] easy directions and a weaker uniaxial term ($K_U$=8.6kJm$^{-3}$) favouring the [110] direction. The symmetry and magnitude of these anisotropy terms is similar to that observed in thin Fe films grown epitaxially onto GaAs (001) substrates [16]. X-ray magnetic circular dichroism (XMCD) measurements (not shown) measured at beamline I06 at the Diamond Light Source reveal that the magnetic moment per Fe atom is not diminished from the value expected for pure Fe.

For electrical transport studies, standard photolithography techniques were used to fabricate a Hall bar of width 45μm with voltage probes separated by 235μm, with the direction of the current along the [110] crystal direction (Fig. 1(c)). Following a similar technique to Ref.5 the chip was bonded onto a piezoelectric transducer capable of producing a uniaxial strain, ε in the layer of order ±10$^{-4}$ at room temperature for applied voltages in the range -40V to +60V. Uniaxial strain was induced along the [010] crystal direction with tensile strain defined as positive along this direction.

Electrical readout of the direction of the magnetization was achieved by measuring the anisotropic magnetoresistance (AMR). To a first approximation, for in plane magnetization, the longitudinal ($\rho_{xx}$) and transverse ($\rho_{xy}$) resistivities are given by $\rho_{xx} = \rho_{av} + \Delta\rho\cos 2\theta$ and $\rho_{xy} = \Delta\rho \sin 2\theta$, where θ is the angle between the magnetization and the current, $\rho_{av}$ is the average of $\rho_{xx}$ when the magnetization is rotated through 360° in the plane, and $\Delta\rho$ is the amplitude of the AMR. From the above relationships it is evident that switching of the magnetization between the [100]/[010] easy axes results in the largest change in the transverse resistivity (also known as planar Hall effect [17]). Figures 2(a) and (b) show the change in $\rho_{xy}$ measured at room temperature as an external magnetic field is applied in the film plane along the [100]/[010] directions for two different voltages ($V_P$) applied to the piezoelectric transducer, corresponding to a tensile or compressive uniaxial strain applied to the Hall bar device. Large changes of $\rho_{xy}$ correspond to large changes of the magnetization orientation as the external magnetic field is swept through zero, indicating that field sweeps along such directions correspond to sweeps along hard magnetic axes. Therefore, the magnetic easy axis can be identified as close to [100] when large compressive strain is applied, and close to [010] for large tensile strain [18]. (See Supplementary Material for the determination of the strain as a function of the voltage.)

The AMR relations were used to deduce the magnetization component along the field direction. An example is shown in Fig. 2(c) for the curve measured with large tensile strain and the magnetic field applied along [100]. The shape of the magnetic hysteresis loop, involving a double step in each field direction, is characteristic of the magnetic reversal behaviour of a sample with biaxial anisotropy [19]. The specific changes in the direction of the magnetization at the switching fields $H_C1$ and $H_C2$ are shown schematically in Fig. 2(d). Similar magnetic hysteresis loops were observed over a range of compressive and tensile induced strains.

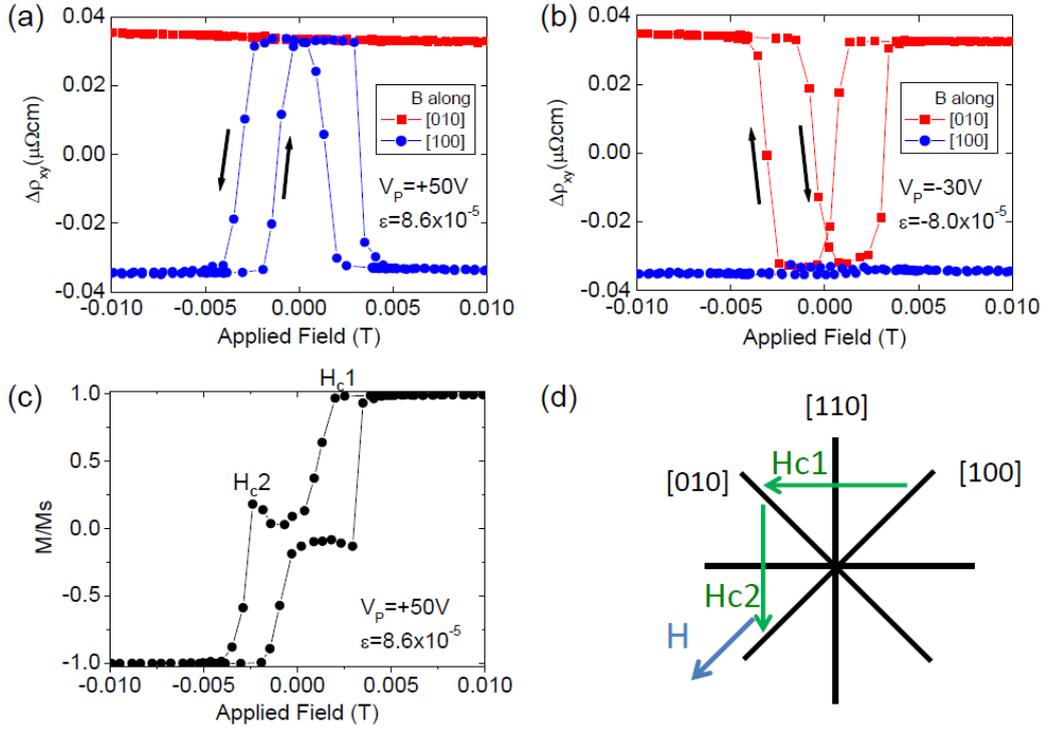

**Figure 2.** (color online) The change in the transverse resistivity, $\Delta\rho_{xy} = \rho_{xy} - \rho_{offset}$ [18] (see Supplemental Material for the determination of $\rho_{offset}$) measured as a function of the magnetic field applied in the plane of the device along the [100]/[010] directions with (a) tensile and (b) compressive strain applied along the direction of the bar. (c) Magnetic hysteresis loop extracted from the data in (a) using the AMR formula for transverse resistivity and field applied along [100]. (d) A schematic diagram showing the switching of the magnetic easy axis at fields $H_C1$ and $H_C2$.

We now demonstrate how the combination of a large magnetostriction and cubic magnetic anisotropy allows us to achieve non-volatile voltage control of the magnetization in our ferromagnet/piezoelectric device in the absence of a magnetic field. Figure 3(a) shows the change in $\rho_{xy}$ measured as the strain was swept from high compressive to high tensile and back several times. The sample was initialised by applying a saturating magnetic field along the [100] direction with a large compressive strain and then placed in µ-metal shielding to minimise any external magnetic fields. $\rho_{xy}$ takes one of two distinct values (α or β in Fig. 3(a)) when zero strain is applied, depending upon the history of the sweep. These values correspond to the magnetization lying along the easy axes close to the [100] and [010] directions respectively. The easy axes at zero strain are rotated approximately 7° towards the [110] due to the small uniaxial component of the anisotropy, $K_U$. The inset shows $\rho_{xy}$ as a function of the voltage applied to the transducer, indicating that our device is non-volatile since the two stable magnetic states also exist at $V_P$=0V. Small differences between the curves plotted against strain or voltage arise due to a small hysteresis present in the strain vs voltage characteristics of the transducer [18].

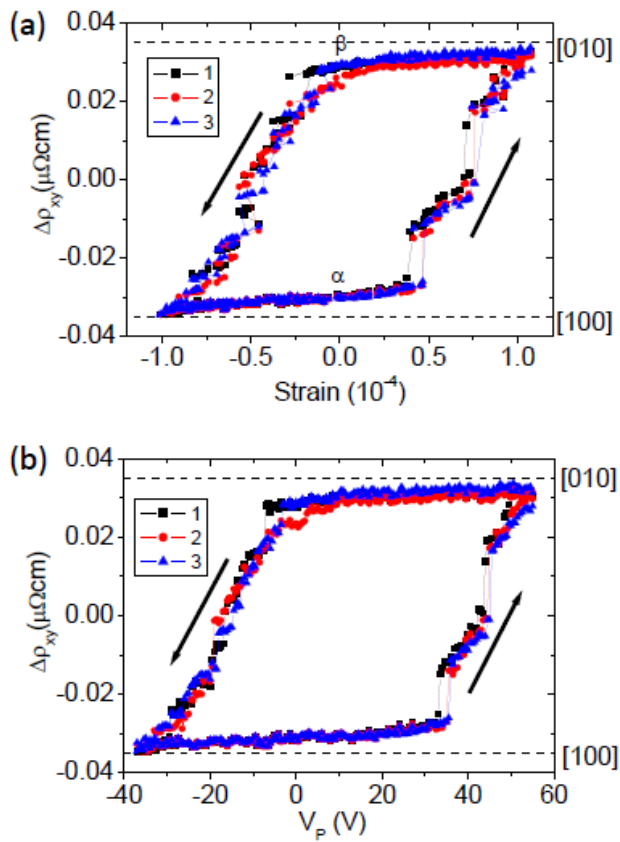

**Figure 3** (Color online) (a) The change in the transverse resistivity, $\Delta\rho_{xy} = \rho_{xy} - \rho_{offset}$ [18] (see Supplemental Material for the determination of $\rho_{offset}$) measured as a function of the applied strain after first saturating in a field of 0.2T and high negative strain. The different symbols represent successive sweeps of the strain, showing the reproducibility of the magnetization switching. The initial loop is not shown. The values for the magnetization pointing along the [100] and [010] directions are indicated by the dashed lines. (b) $\Delta\rho_{xy}$ as a function of $V_P$.

The non-volatile switching of the magnetization, and the role of domain walls in the magnetization reversal process are clearly demonstrated by spatially resolved magneto-optical Kerr effect (MOKE) measurements on the same section of the Hall bar as was used for the measurement of $\rho_{xy}$. Magnetic domains were imaged using a commercial wide-field magneto-optical Kerr microscope. Illumination of the sample at oblique incidence allows sensitivity to the in-plane moments due to the longitudinal Kerr effect. The opposite dark or light contrast observed in the central region of the Hall bar in Figs. 4(a) (ii) and (iv) confirm that the direction of the magnetization at zero strain is dependent upon the history of the strain sweep and is set by either a high tensile or compressive strain. Panels (i),(iii) and (v) show that, at the highest tensile and compressive strains used in our experiment the magnetization in some regions of the Hall bar is not fully reversed. These vestigial domains may be the result of domain wall pinning sites that could arise due to local inhomogeneity in the magnetic anisotropy as a result of imperfections in the lithography or local variations of the strain. The vestigial domains seed the magnetization reversal with domain walls moving over distances of tens of microns until the switched domains coalesce to fill the entire central region of the Hall bar device. Figure 4(b) shows more detailed images of the domain wall motion as a function of the induced strain during this process. These images are consistent with the multistep switching observed in Fig. 3(a) and indicate that the magnetization switching involves domain wall processes with the domain walls overcoming multiple pinning sites. We note that the strain-induced movement of domain walls presented here occur over distances that are at least an order of magnitude larger than previous observations of domain walls in magnetostrictive materials [20,21]

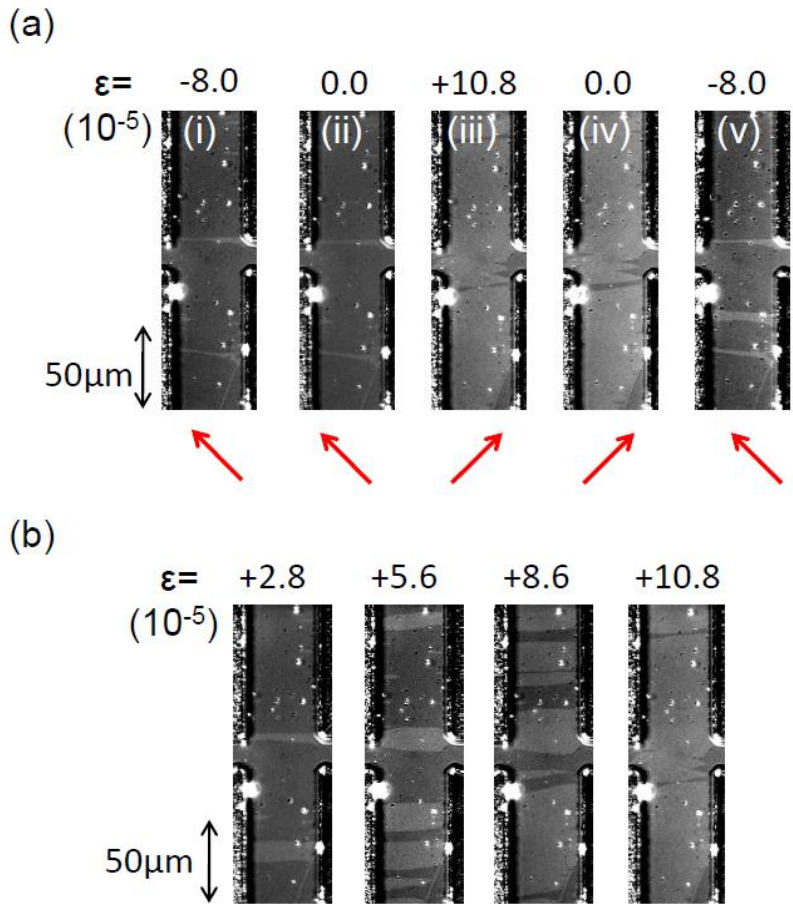

**Figure 4.** (Color online) (a) MOKE images showing the section of the Hall bar used for the measurements of $\rho_{xy}$ for different applied strain ($\varepsilon$). The contrast represents the component of the magnetization pointing across the bar. The arrows beneath each image indicate the direction of the magnetization in the region between the voltage probes. (b) The evolution of the magnetic configuration by domain wall motion as a function of the induced strain. The sequence represents the transition between panels (ii) and (iii) in Fig. 4(a).

In summary, we have demonstrated that thin epitaxial $Fe_{81}Ga_{19}$ films possess the desirable combination of cubic magnetocrystalline anisotropy and large magnetostriction. This has been used to demonstrate non-volatile switching of the magnetization by a voltage induced uniaxial strain. This was achieved at room temperature and was shown to be reproducible without the application of an external magnetic field. In our device the magnetic switching occurs with the application of voltages in the range -40V to +60V. However, by scaling such devices down to sub-micron sizes and integrating piezoelectric and magnetostrictive layers, fast magnetization switching will be achievable with the application of only a few volts. The new functionality demonstrated in this work could be combined with intentionally designed pinning sites and shape anisotropy gradients to allow reversible and irreversible control of domain walls via voltage induced strain. Such devices will be suitable for applications in information storage and logical processing.

The authors acknowledge financial support from EPSRC grant number EP/H003487/1 and EU grant No. NAMASTE 214499. We are grateful for helpful discussions with Dr Alexander Shick, Dr Jan Zemen and Prof. Tomas Jungwirth.

# Non-volatile voltage control of magnetization and magnetic domain walls in magnetostrictive epitaxial thin films.

## Supplementary Material

**Measurement of the strain**

The piezoelectric transducer used in our device is expected to produce a strain of order $10^{-3}$ over the voltage range studied. Not all of this strain will be transmitted to the $Fe_{81}Ga_{19}$ film due to the elastic properties of the GaAs substrate and the glue used to bond the chip to the transducer. To characterise the strain transmitted to the top layer we fabricated a $Ni_{80}Cr_{20}$ strain gauge in a resistance bridge geometry on a GaAs substrate. The GaAs chip was then thinned and glued to a piezoelectric transducer using identical processing steps as the first device. The strain, determined from the measurement of the resistance bridge as a function of the voltage applied to the piezoelectric transducer is shown in Figure 1. Also shown is the strain determined from the fractional change in the resistance of the Hall bar when measured in a saturating magnetic field of 0.5T to suppress rotations of the magnetization and therefore any effects due to magnetostriction. The close agreement between the two curves indicates that the change in the resistance of the bar represents a good measure of the strain and this was the method used to quantify the strain for subsequent analysis. Differences between the curves may arise from variations in the properties of the piezoelectric transducers, or may be due to the extent to which the strain is transmitted in each device. Results from both methods suggest that the transmitted strain is an order of magnitude lower than that generated by the transducer alone. The transducers used in our study are from the same supplier as those studied by Shayegan et al [1]. The shape of the curves, including the presence of hysteresis, is consistent with the behaviour observed in that study.

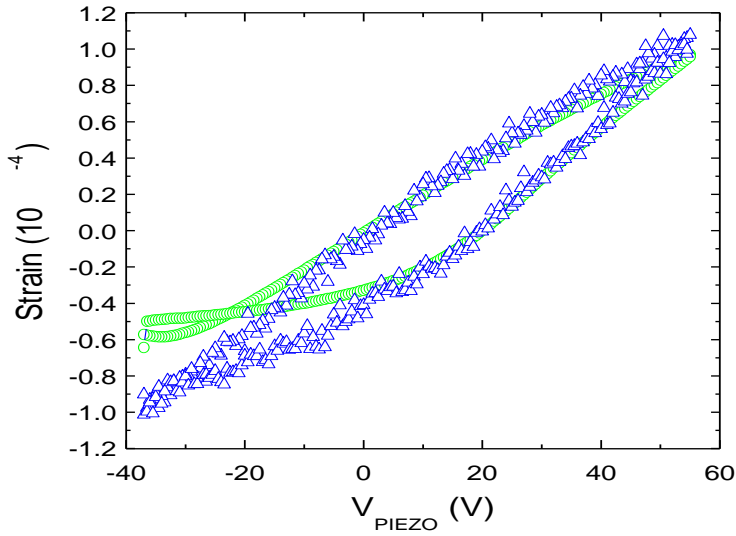

**Figure 1. (Color online)** The uniaxial strain along the direction of the piezoelectric stressor as a function of the voltage applied. The strain was determined at T=300K using a $Ni_{80}Cr_{20}$ resistance bridge (green circles) and using the change in the longitudinal resistance of the $Fe_{81}Ga_{19}$ Hall bar in a saturating magnetic field of 0.5T applied along the direction of the current (blue triangles).

**Anisotropic magnetoresistance measurements**

The measured value of $\rho_{xy}$ includes an offset, $\rho_{offset}$ which is independent of the magnetic field. This arises from the imperfect alignment of the voltage probes, or inhomogeneities in the resistivity of the bar. In the present study any inhomogeneity in strain will contribute to this offset which can therefore be strain dependent. $\rho_{offset}$ is approximately 1 part in $10^4$ of the longitudinal resistance and has a weak dependence on the strain. To determine the strain dependence we measured $\rho_{xy}$ as a function of strain with a saturating magnetic field along the [100] and [010] directions. This is shown in Figure 2. The dependence of the offset on strain is approximately linear and the slope of each line, $d\rho_{xy}/d\varepsilon$ =156.8µΩcm for both directions of the magnetic field. We define $\rho_{offset}$ as the average of the curves along the [100] and [010] directions. This has been subtracted from the measured data in Figures 2 and 3 to show $\Delta\rho_{xy}$

= $\rho_{xy} - \rho_{offset}$. After subtraction of this offset the variations in the value of $\rho_{xy}$ in Figures 2 and 3 arise solely from changes in the direction of the magnetization.

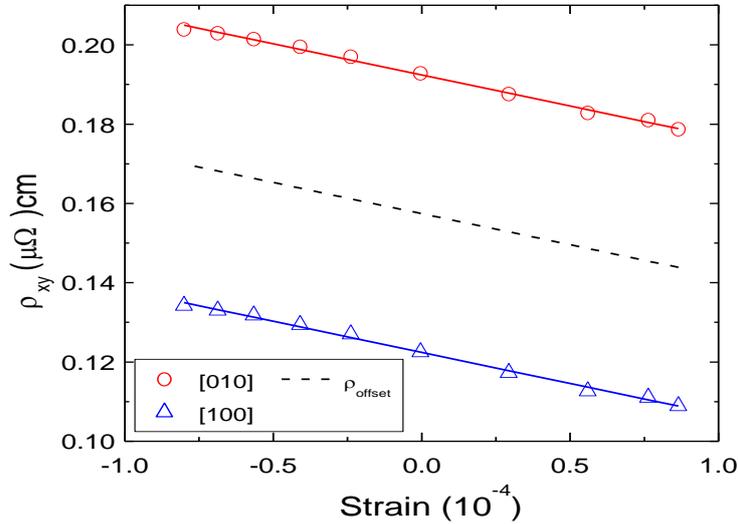

**Figure 2. (Color online)** The graph shows the measured values of $\rho_{xy}$ (symbols) as a function of the strain with a saturating magnetic field applied along the [100] and [010] directions. Solid lines represent a linear fit to the data. $\rho_{offset}$ is represented by the dashed line and is defined as the average of the lines for the [100] and [010] directions.

[1] M. Shayegan, K. Karrai, Y.P. Shkolnikov, K. Vakili, E.P. De Poortere, S. Manus, Appl. Phys. Lett. **83**, 5235 (2003).